\documentclass
{IEEEtran} 
\usepackage{amsmath,amssymb,verbatim,rotating,graphicx} 

\newtheorem{theorem}{Theorem}
\newtheorem{lemma}{Lemma}
\newtheorem{corollary}{Corollary}
\newtheorem{definition}{Definition}
\newcommand{\enproof}{\hfill $\Box$ \vspace*{1ex}}

\newcommand{\ssi}{^{(i)}} 
\newcommand{\mbssi}{\mbox{}^{(i)}}

\newcommand{\mymathbb}[1]{{\mathbb{#1}}} 
\newcommand{\mymathsf}[1]{{\mathsf{#1}}} 

\newcommand{\dmn}{q} 

\newcommand{\ssf}{\mymathsf{f}}
\newcommand{\myF}{{\mymathbb{F}_{\dmn}}} 
\newcommand{\myFnoarg}{{\mymathbb{F}}}
\newcommand{\myFk}{{\mymathbb{F}_{\dmn^k}}}

\newcommand{\sQ}{\mymathsf{Q}}

\renewcommand{\phi}{\varphi} 
\renewcommand{\subset}{\subseteq}
\renewcommand{\tilde}{\widetilde}
\renewcommand{\hat}{\widehat}
\renewcommand{\bar}{\overline}

\newcommand{\SINT}{\mymathbb{Z}}

\newcommand{\transp}{\mbox{}^{\rm t}}

\newcommand{\tracenos}{{\rm Tr}\,}
\newcommand{\trace}{{\rm Tr}_{\myFk/\myF}\,}

\newcommand{\crd}[1]{|#1|}

\newcommand{\dpr}[2]{#1 \cdot #2}

\newcommand{\spn}{\mymathsf{span}\,}

\newcommand{\Ccl}{C}
\newcommand{\Rcl}{R_c} 
\newcommand{\rcl}{r_{\rm c}} 
\newcommand{\rqt}{r_{\rm q}}

\newcommand{\myFpower}[1]{\mymathbb{F}_{\dmn}^{#1}}

\newcommand{\myFpowernoarg}[1]{\mymathbb{F}^{#1}}

\newcommand{\Qpl}[1]{\sQ^+} 
\newcommand{\Qmi}[1]{\sQ^-}
\newcommand{\vart}{b} 

\newcommand{\Bp}[1]{G}

\newcommand{\zrvb}{{\bf 0}}

\newcommand{\Jgood}{\tilde{J}}

\newcommand{\crI}{J} 

\newcommand{\Jof}[1]{\prm(\Jgood)}

\newcommand{\prm}{\pi}

\newcommand{\chooses}[2]{\Big( \begin{array}{c} #1 \\[-1ex] #2 \end{array} \Big)}

\newcommand{\kperp}{\perp} 
\newcommand{\embone}{\pi_1}
\newcommand{\embtwo}{\pi_2}
\newcommand{\Cone}{L_1}
\newcommand{\Ctwo}{L_2}
\newcommand{\tnum}{M}
\newcommand{\intint}[2]{[#1,#2]_{\SINT}} 

\newcommand{\degF}{k} 
\newcommand{\intk}{l} 
\newcommand{\MatF}{\Phi}
\newcommand{\varKgen}{K}
\newcommand{\bsa}{\mymathsf{a}}
\newcommand{\bsb}{\mymathsf{b}}
\newcommand{\bab}[1]{\begin{matrix} | \\ #1 \\ | \end{matrix}}
\newcommand{\myFkpower}[1]{\myFnoarg_{\dmn^\degF}^{#1}}

\newcommand{\MatFa}{\MatF_{\bsa}}

\newcommand{\subgrp}{\le}

\newcommand{\CSone}{C_1}
\newcommand{\CStwo}{C_2^{\perp}} 
\newcommand{\CStwp}{C_2}
\newcommand{\Noa}{N_{\rm o}}
\newcommand{\Koa}{K_{\rm o}}
\newcommand{\Roa}{R_{\rm o}}

\newcommand{\Pe}{P_{\rm e}}
\newcommand{\Peone}{P_{{\rm e},1}}
\newcommand{\Petwo}{P_{{\rm e},2}}
\newcommand{\Pej}{P_{{\rm e},j}}
\newcommand{\Pein}{P}
\newcommand{\Peinj}{P_j}

\title{Concatenated Conjugate Codes}

\author{Mitsuru Hamada, {\em Member, IEEE}
\footnote{The author is with 
Research Center for Quantum Information Science,
               Tamagawa University Research Institute,
6-1-1 Tamagawa-gakuen, Machida, Tokyo 194-8610, Japan,
and with PRESTO, Japan Science and Technology Agency, 
4-1-8 Honcho, Kawaguchi, Saitama, Japan.
E-mail: {\tt mitsuru@ieee.org}.}
}

\begin{document}

\maketitle

\begin{abstract}
A conjugate code pair is defined as a pair of linear codes
either of which contains the dual of the other.
A conjugate code pair represents the essential structure
of the corresponding Calderbank-Shor-Steane (CSS) quantum code.
It is known that conjugate code pairs are applicable to
(quantum) cryptography.
We give a construction method for efficiently decodable conjugate code pairs. 
\end{abstract}


\section{Introduction \label{ss:intro}}

A conjugate code pair is a pair of linear codes $(C_1,C_2)$
satisfying the condition $C_2^{\perp} \subset C_1$,
where $C^{\perp}$ denotes the dual of $C$.
This paper 
treats the issue of constructing
a conjugate code pair $(C_1,C_2)$ such that either $C_1$ or $C_2$
(more precisely, either $C_1/C_2^{\perp}$ or
$C_2/C_1^{\perp}$; see Section~\ref{ss:qc_cc}) 
are efficiently decodable.
Namely, 
we give a construction method for efficiently decodable conjugate code pairs. 
Motivations for constructing such pairs are given in 
\cite{ShorPreskill00,mayers02,hamada03s,hamada06s,hamada05qc} and briefly described below.

In the past decades, great efforts have been made to extend
information theory and its ramifications to quantum theoretical
settings. 
In particular, after a proof of the 
`unconditional' security of a 
quantum key distribution (QKD) protocol~\cite{BennettBrassard84}
was given~\cite{mayers01acm},
it was observed~\cite{ShorPreskill00} that
the structure of Calderbank-Shor-Steane 
(CSS) codes~\cite{CalderbankShor96,steane96a} 
had been used implicitly in the QKD protocol. 
Moreover, it was argued~\cite{ShorPreskill00} that the security of the QKD
protocol could be proved by bounding the fidelity (a performance measure, 
which parallels the probability of successful decoding) of 
CSS codes underlying the protocol.  

CSS codes are a class of algebraic quantum error-correcting codes, called
symplectic codes, or stabilizer codes~\cite{crss97,crss98,gottesman96}. 
The term conjugate code pairs or conjugate codes~\cite{hamada06s} is
almost a synonym for CSS codes if one forgets about quantum mechanical
operations for encoding or decoding and pays attention only to
what can be done in the coding theorists' universe of finite fields.
Namely, a CSS code is specified by a conjugate code pair $(C_1,C_2)$.%
%
\footnote{%
The bridge between the coding theorists universe, 
the vector space $\myFpower{2n}$ over a finite field $\myF$, 
and quantum mechanical worlds
that are represented by Hilbert spaces 
is Weyl's projective representation $N$ of $\myFpower{2n}$,
which maps a vector in $\myFpower{2n}$ to a unitary operator
on a $\dmn^{n}$-dimensional Hilbert space~\cite{weyl28}.
In fact, a symplectic code is a simultaneous eigenspace of
a set of commuting operators that can be written as
$N(S)$ or $N_S$, the image of $S \subset \myFpower{2n}$,
and a CSS code is such that $S$ is specified by a conjugate code pair 
$(C_1,C_2)$ via $S=\{ [u,v] \mid u\in C_1^{\perp},v\in C_2^{\perp} \}$ in the notation of \cite{hamada05qc,hamada06s,hamada03s}.}

It is known that if codes $C_1$ and $C_2$ are both good,
the CSS quantum code specified by $C_1$ and $C_2$ is good, and hence,
the cryptographic code or QKD protocol resulting from $(C_1,C_2)$ is good in view of security and
reliability (probability of successful decoding).
In this context, either $C_1$ or $C_2$ 
should be efficiently decodable because only one of the two codes is used 
for transmission of secret data.

It may be interesting that only the `structure of' CSS codes is 
used in the QKD protocol above mentioned. In other words,
what is used in the QKD protocol is not a quantum code but
a reduced form of a CSS code, and this reduced form is a linear
error-correcting code.
More precisely,  this is a quotient code~\cite{hamada05qc} 
of the form $C_1/C_2^{\perp}$, 
which will be explained shortly.
This can be viewed as an error-correcting code that can protect
information from eavesdroppers.
Quotient codes 
fall in the class of coding systems devised in a similar but classical context
in \cite{wyner75}, though
we have arrived at this notion through a different path, i.e.,
through explorations on quantum cryptography~\cite{BennettBrassard84,ShorPreskill00,mayers01acm,hamada03s}.
(The adjective `classical' will sometimes refer to
not being quantum theoretical.)
We remark that as is implicit in \cite{hamada03s}
and explicit in \cite{hamada06s},
quotient codes 
can be used as cryptographic codes that are more general than QKD schemes.
(General cryptographic codes allow direct encoding of secret data, whereas
the aim of key distribution is to share a random string between 
remote sites.)

In \cite{CalderbankShor96,hamada03s},
the existence of good 
CSS codes was proved by random coding. 
In particular, 
the rate $1-2h(p)$,
where $h$ denotes the binary entropy function,
was called the Shannon rate in \cite{ShorPreskill00}
and proved achievable in \cite{hamada03s}.
However, 
these codes do not have a rich 
structure that allows efficient decoding.
In this paper, we consider the issue of constructing efficiently
decodable conjugate codes.
Our approach is that of concatenated codes~\cite{forney},
by which we establish that the rate $1-2h(p)$ is achievable 
with codes of polynomial decoding complexity.

Besides applications to cryptography, our construction
gives quantum error-correcting codes superior to those known~\cite{hamada06expl,hamada06md,hamada06itw}.

We remark another major approach, i.e., 
that of low density parity check codes had already been
taken to construct CSS codes~\cite{MacKayMM04}.
However, the present work is different from \cite{MacKayMM04} in 
that the decoding error probability is evaluated without approximation
or resort to simulation. 

This paper is organized as follows.
In Section~\ref{ss:qc_cc}, we introduce quotient codes and
conjugate codes. In Section~\ref{ss:ccc},
concatenated conjugate codes are defined.
In Sections~\ref{ss:dec} and \ref{ss:synd_dec}, 
methods for decoding are described.
The performance of concatenated conjugate codes is evaluated
in Section~\ref{ss:performance}.
Section~\ref{ss:dis_rem} contains discussions and remarks.
Section~\ref{ss:sum} contains a summary.
An appendix is given for proving a fundamental lemma, on which
our construction is based.

\section{Quotient Codes and Conjugate Codes \label{ss:qc_cc}}

We fix some notation.
The set of consecutive integers 
$\{ l,l+1,\dots,m \}$ is denoted by $\intint{l}{m}$. 
We write $B \subgrp C$,
or $C \ge B$, 
if $B$ is a subgroup of an additive group $C$.
We use the dot product defined by
$ 
\dpr{(x_1,\dots,x_n)}{(y_1,\dots,y_n)}=\sum_{i=1}^{n} x_iy_i
$ 
on $\myFpowernoarg{n}$, where $\myFnoarg$ is a finite field.
For a 
subset $\Ccl$ of $\myFpowernoarg{n}$, 
$C^{\perp}$ denotes $\{ y \in\myFpowernoarg{n} \mid \forall x\in C, \ \dpr{x}{y}=0 \}$.
A subset $\Ccl$ of $\myFpowernoarg{n}$ is called an $[n,k]$ code if
$k=\log_{\crd{\myFnoarg}} \crd{C}$.
As usual, $\lfloor a \rfloor$ denotes the largest
integer $a'$ with $a'\le a$, and $\lceil a \rceil = - \lfloor - a \rfloor$.
The transpose of a matrix $A$ is denoted by $A\transp$.

First, we explain quotient codes introduced in \cite{hamada05qc}.
The aim of \cite{hamada05qc} was to exhibit the essence,
at least, for algebraic coding theorists,
of algebraic quantum coding, 
and this attitude was retained to
introduce the notion of conjugate codes~\cite{hamada06s}.
Throughout, 
we fix a finite field $\myF$ of $q$ elements.
We will construct codes over $\myF$.

A {\em quotient code}\/ of length $n$ over $\myF$ is an additive quotient group
$C/B$ with $B\le C\le\myFpower{n}$.
In the scenario of quotient codes in \cite{hamada05qc}, 
the sender encodes a message into a member $c$ of $C/B$, 
chooses a word in $c$
according to some probability 
distribution on $c$, and then sends it through the channel.
Clearly, if $C$ is a $J$-correcting ($J\subset \myFpower{n}$) 
in the ordinary sense,
$C/B$ is $(J+B)$-correcting (since adding a word in $B$ to the `code-coset'
$c$ does not change it).
The (information) rate of the quotient code $C/B$ is defined as
$n^{-1}\log_{q} |C|/|B|$.

We mean by an $[[n,k]]$ {\em conjugate (complementary) code pair},\/
or CSS code pair, over $\myF$
a pair $(C_1,C_2)$ consisting of
an $[n,k_1]$ linear code $\CSone$ and an
$[n,k_2]$ linear code $C_2$ 
satisfying
%
\begin{equation}\label{eq:css_cond}
\CStwo \subgrp \CSone, 
\end{equation}
which condition is equivalent to
$\CSone^{\perp} \subgrp \CStwp$,
and
\begin{equation}\label{eq:css_k}
k=k_1+k_2-n.
\end{equation}
If $C_1$ and $C_2$ satisfy (\ref{eq:css_cond}), 
the quotient codes $C_1/C_2^{\perp}$ and $C_2/C_1^{\perp}$ 
are said to be conjugate.
The number $k/n$ is called the (information) rate of 
the conjugate code pair $(C_1,C_2)$, and equals
that of $C_1/C_2^{\perp}$ and that of $C_2/C_1^{\perp}$.

The condition (\ref{eq:css_cond}) is equivalent to
that $C_1^{\perp}$ and $C_2^{\perp}$ are perpendicular to each other.
Here, with two codes $C$ and $C'$ given, we say $C$ is perpendicular to $C'$ 
and write
\[
C \perp C'
\]
if $\dpr{x}{y}=0$ for any $x \in C$ and $y\in C'$.
Note that $C \perp C'$ 
if and only if (iff) $C'\le C^{\perp}$, or equivalently, 
iff $C\le C'\mbox{}^{\perp}$.

The goal 
is to find a conjugate code pair $(C_1,C_2)$ such that
both $C_1/C_2^{\perp}$ and $C_2/C_1^{\perp}$ 
have good performance. If the linear codes $C_1$ and $C_2$ both have
good performance, so do $C_1/C_2^{\perp}$ and $C_2/C_1^{\perp}$. 
Hence, a conjugate code pair $(C_1,C_2)$ with good (not necessarily a technical term) 
$C_1$ and $C_2$ is also desirable.
The details may be found in \cite{hamada06s,hamada05qc} or in the other 
literature on CSS codes.

\section{Concatenated Conjugate Codes \label{ss:ccc}}

Forney~\cite{forney} invented a method for
creating error-correcting codes of relatively large lengths
by concatenating shorter codes.  
We bring Forney's idea into our issue of constructing
long conjugate codes. 

\begin{lemma}\label{prop:gg'}
Assume $(C_1, C_2)$ is a conjugate code pair having the parameters as above, 
and 
\begin{equation*}
C_1=C_2^{\perp}+\spn\{ g_1,\dots,g_k \}.
\end{equation*}
Then, we can find vectors $g'_1,\dots,g'_k$ such that
\begin{equation*}
C_2=C_1^{\perp}+\spn\{ g'_1,\dots,g'_k \}
\end{equation*}
and
\begin{equation*}
\dpr{g_l}{g'_m} = \delta_{lm}
\end{equation*}
where $\delta_{lm}$ is the Kronecker delta.
\end{lemma}

\begin{figure} 
\begin{center}
\includegraphics[scale=0.8]{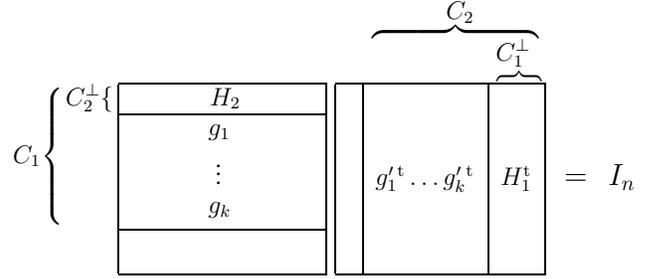} 
\caption{A basic structure of an $[[n,k]]$ conjugate code pair.
\label{fig:css}}
\end{center}
\end{figure}

{\em Proof.}\/ 
We see this from Fig.~\ref{fig:css}.
In fact, 
if $C_1=C_2^{\perp}+\spn\{ g_1,\dots,g_k \} \le \myFpower{n}$ 
and $H_2$ is a full-rank parity check matrix
of $C_2$, we have an invertible matrix, $A$, as depicted at the left-most
position of Fig.~\ref{fig:css}.
Of course, we have its inverse $A^{-1}$, which is depicted next to $A$
in the figure.
Write $g'_1\transp,\ldots,g'_k\transp$ for the $(n-k_2+1)$-th to 
$k_1$-th columns of $A^{-1}$.
Then, we see that $\dpr{g_l}{g'_m} = \delta_{lm}$ and the last $n-k_1$ columns 
of the second matrix are
perpendicular to the $[n,k_1]$ code $C_1$. 
\mbox{} \enproof 

Let $(C_1\ssi,C_2\ssi)$, $i\in[1,N]_\SINT$, be $[[n\ssi,k]]$ conjugate code pairs over $\myF$,
where $C_1$ and $C_2$ are $[n\ssi,k_1\ssi]$ and $[n\ssi,k_2\ssi]$ codes, respectively,
with
\[
k=k_1\ssi+k_2\ssi-n\ssi, \quad i\in[1,N]_\SINT.
\]

Assume $g_l\ssi$ and $g'_l\mbssi$, $l\in[1,k]_\SINT$, satisfy
the conditions in Lemma~\ref{prop:gg'}. In particular,
\begin{equation}\label{eq:gg'}
\dpr{g_l\ssi}{g'_m\mbssi} = \delta_{lm}.
\end{equation}


The field $\myFk$ is an $\myF$-linear vector space, 
and we can take bases $(\beta_j)_{j=1}^{k}$ and
$(\beta'_j)_{j=1}^{k}$ 
that are dual to each other 
with respect to the $\myF$-bilinear form (e.g., \cite{LidlNied,sloan})
defined by
\begin{eqnarray*}
\ssf &:& \myFk\times\myFk \to \myF,\\
&& (x,y) \mapsto \trace xy.
\end{eqnarray*}
In particular, $\ssf(\beta_l,\beta'_m)=\delta_{lm}$.

Now we can define a pair of maps 
that preserve the bilinear form (inner product) as follows. Let
\begin{eqnarray*}
\embone\ssi &:& \myFk \to \spn \{ g_1\ssi,\dots,g_k\ssi \} \simeq C_1\ssi/C_2^{\perp}\mbssi,\\
&& \sum_j x_j \beta_j \mapsto \sum_j x_j g_j\ssi,
\end{eqnarray*}
and $\tilde{C_1\ssi}$ denote $\spn \{ g_1\ssi,\dots,g_k\ssi \}$.
Let
$\bigoplus_{i=1}^N y\ssi$ denote the concatenated vector $(y_1\ssi \cdots y_N\ssi)\in\myFpower{\sum_{i=1}^N n\ssi}$ for $y\ssi=(y_1\ssi,\cdots,y_{n\ssi}\ssi)\in\myFpower{n\ssi}$, 
$i\in\intint{1}{N}$,
and $\bigoplus_{i=1}^N A\ssi$ denote the set of those vectors
$\bigoplus_{i=1}^N y\ssi$ with $y\ssi\in A\ssi\le \myFpower{n\ssi}$, 
$i\in\intint{1}{N}$.

We can compose a larger map applying $\embone\ssi$ to 
the $i$-th coordinate of a vector in $\myFkpower{N}$:
\begin{eqnarray*}
\embone &:& \myFkpower{N} \to \bigoplus_{i=1}^N \tilde{C_1\ssi},\\
&& \bigoplus_{i=1}^N \sum_j x_j\ssi \beta_j \mapsto \bigoplus_{i=1}^N \sum_j x_j\ssi g_j\ssi.
\end{eqnarray*}
Similarly, we define 
\begin{eqnarray*}
\embtwo &:& \myFkpower{N} \to \bigoplus_{i=1}^N \tilde{C_2\ssi},\\
&& \bigoplus_{i=1}^N \sum_j x_j\ssi \beta'_j \mapsto \bigoplus_{i=1}^N \sum_j x_j\ssi g'_j\mbssi.
\end{eqnarray*}

Then, for $x=(x^{(1)},\dots,x^{(N)})$ and 
$y=(y^{(1)},\dots,y^{(N)})$ with
\[
x\ssi=\sum_j x_j\ssi \beta_j \quad \mbox{and} 
\quad y\ssi=\sum_j y_j\ssi \beta'_j,
\]
we have
\begin{equation}
\trace x \cdot y = \dpr{\embone(x)}{\embtwo(y)}. \label{eq:preserveip}
\end{equation}
This can be seen by noting
\begin{eqnarray*}
\lefteqn{ \trace x\ssi y\ssi } \nonumber\\
&=& \ssf(x\ssi, y\ssi) \nonumber\\
&=& \ssf(\sum_j x_j\ssi \beta_j,\sum_j y_j\ssi \beta'_j )\nonumber\\
&=& \sum_{j=1}^k x_j\ssi y_j\ssi \nonumber\\
&=& \dpr{\left(\sum_j x_j\ssi g_j\ssi\right)}
{ \left(\sum_j y_j\ssi g'_j\mbssi\right)} \nonumber\\
&=&\dpr{\embone\ssi(x\ssi )}{\embtwo\ssi(y\ssi)}
\end{eqnarray*}
and taking summations of the end sides over $i\in\intint{1}{N}$.

\begin{definition}\label{def:1}
The concatenation (or concatenated conjugate code pair made) 
of conjugate code pairs $(C_1\ssi,C_2\ssi)$ over $\myF$,
$i\in\intint{1}{N}$,
and 
an $[[N,K]]$ conjugate code pair $(D_1,D_2)$ over $\myFk$
is the $[[\sum_{i=1}^{N} n\ssi,kK]]$ conjugate code pair
\[
(\embone(D_1)+\overline{C_2^{\perp}}, [\embone(D_2^{\kperp})+\overline{C_2^{\perp}}]^{\perp})
\]
over $\myF$,
where 
\[
\overline{C_m^{\perp}}=\bigoplus_{i=1}^N C_m^{\perp}\mbssi, \quad m=1,2.
\]
\end{definition}

If $(C_1\ssi,C_2\ssi)$ is identical to a fixed $[[n,k]]$ conjugate code 
pair $(C_1,C_2)$,
it is called the concatenation of $(C_1,C_2)$ and $(D_1,D_2)$.
It is an $[[nN,kK]]$ conjugate code pair.
The codes $C_1\ssi,C_2\ssi$ are sometimes called inner codes,
and $D_1,D_2$ outer codes.

\begin{theorem}\label{th:duals_ccc}
\[
[\embone(D_2^{\kperp})+\overline{C_2^{\perp}}]^{\perp} = \embtwo(D_2)+\overline{C_1^{\perp}},
\]
\[
[\embtwo(D_1^{\kperp})+\overline{C_1^{\perp}}]^{\perp} = \embone(D_1)+\overline{C_2^{\perp}}.
\]
\mbox{}
\end{theorem}

\begin{corollary}
The concatenated conjugate code pair in Definition~\ref{def:1} 
can be written as
\[
(\embone(D_1)+\overline{C_2^{\perp}}, \embtwo(D_2)+\overline{C_1^{\perp}}).
\]
\mbox{}
\end{corollary}

{\em Proof.}\/ It is enough to prove the second equality by virtue of 
the symmetry. 
First, we show
\begin{equation}\label{eq:2prove}
[\embtwo(D_1^{\kperp})+\overline{C_1^{\perp}}]^{\perp} \ge \embone(D_1)+\overline{C_2^{\perp}},
\end{equation}
which is equivalent to 
\[
\embone(D_1)+\overline{C_2^{\perp}} \perp
\embtwo(D_1^{\kperp})+\overline{C_1^{\perp}}.
\]
The code $\embone(D_1)$ is perpendicular to $\embtwo(D_1^{\kperp})$
by (\ref{eq:preserveip}), 
and to $\overline{C_1^{\perp}}$ trivially.
Similarly, $\overline{C_2^{\perp}}$ is perpendicular
to $\embtwo(D_1^{\kperp})$.
By the CSS property (\ref{eq:css_cond}), $C_2\ssi\mbox{}^{\perp}$ and
$C_1\ssi\mbox{}^{\perp}$ are perpendicular to each other,
and hence, $\overline{C_2^{\perp}}$ is perpendicular to
$\overline{C_1^{\perp}}$.

Thus, we have (\ref{eq:2prove}).
Since 
$\dim_{\myF}[\embtwo(D_1^{\kperp})+\overline{C_1^{\perp}}]+ \dim_{\myF} [\embone(D_1)+\overline{C_2^{\perp}}]=\sum_{i=1}^{N} n\ssi$,
we have the lemma,  and hence, the corollary.
\enproof


Note that a generator matrix of 
$\embtwo(D_1^{\perp})+\overline{C_1^{\perp}}$ over $\myF$
has the form 
\begin{equation}\label{eq:H4cat}
\begin{bmatrix}
H_1^{(1)} &O & \dots & O \\
O & H_1^{(2)} & & O \\
\vdots &  &  \begin{rotate}{45}$\vdots$\end{rotate}  &   \\
O & O & & H_1^{(N)} \\
G'_{1,1} & G'_{1,2} & \cdots & G'_{1,N} \\
\vdots &  \vdots  &        & \vdots \\
G'_{\tnum,1} & G'_{\tnum,2} & \cdots & G'_{\tnum,N}
\end{bmatrix}
\end{equation}
where $H_1\ssi$ is a parity check matrix of $C_1\ssi$, 
$O$ is the zero matrix (whose size may vary from place to place),
$\tnum=N-K_1$ ($K_1$ is the dimension of $D_1$), 
and for each $(i,j)$, $G'_{j,i}$ is
an $n\ssi \times k$ matrix whose rows are spanned by $g'_l\mbssi$.
Hence, by Theorem~\ref{th:duals_ccc}, (\ref{eq:H4cat}) is a parity check matrix
of $\embone(D_1)+\overline{C_2^{\perp}}$.

\section{Decoding Strategy for Concatenated Conjugate Codes \label{ss:dec}}

We investigate correctable errors of the concatenated quotient codes
$\Cone/\Ctwo^{\perp}$,
where
$\Cone=\embone(D_1)+\overline{C_2^{\perp}}$ 
and $\Ctwo=[\embone(D_2^{\kperp})+\overline{C_2^{\perp}}]^{\perp} =\embtwo(D_2)+\overline{C_1^{\perp}}$,
under the scenario of quotient codes described in Section~\ref{ss:qc_cc}
or in \cite{hamada05qc}.
This is a half of the conjugate code pair 
$(\Cone/\Ctwo^{\perp},\Ctwo/\Cone^{\perp})$,
and the other half, having the same form, can be treated similarly.

We remark that in known applications of conjugate codes, i.e.,
for CSS quantum codes and cryptographic codes as in 
\cite{ShorPreskill00,hamada03s,hamada06s},
the decoding should be a {\em syndrome decoding}, which consists of
measuring the syndrome, estimating the error pattern, and 
canceling the effect of the error.

We decode the code in the following two stages.
\begin{enumerate}
\item For each of the inner quotient codes $C_1\ssi/C_2\ssi$, we perform
a syndrome decoding (as described in Sections~2 and 3 of \cite{hamada05qc} for preciseness).
\item For the outer code $D_1$, we perform an efficient decoding such as
bounded distance decoding. 
\end{enumerate}

For efficient decoding, the outer code
$D_1$ should allow a decoding algorithm of polynomial complexity in $N$.
Assume $n\ssi=n$ for all $i$ for simplicity.
Then, if $N \ge q^{\tau k}$ and $k/n \to r$ as $n\to \infty$,
where $\tau>0$ and $r \ge 0$ are constants,
the concatenated conjugate codes $\Cone/\Ctwo^{\perp}$ 
can be decoded with polynomial complexity in $N$,
and hence in the overall code-length $nN$.
Generalized Reed-Solomon (GRS) codes~\cite{sloan} are examples of such codes.

Now assume the sender sent a word $x \in (\myFpower{n})^N$,
$x$ suffered an additive error $e=(e_1,\dots,e_N)\in (\myFpower{n})^N$,
and the receiver received a word $y=x+e\in (\myFpower{n})^N$.
Using the upper half of 
the parity check matrix in (\ref{eq:H4cat}), where $H_1\ssi$ are involved,
the receiver decodes the inner quotient codes.
Namely, receiver estimates $e_i$, 
and subtract $\hat{e}=(\hat{e}_1,\ldots, \hat{e}_N)$ from $y$,
where $\hat{e}_i$ is the estimate of $e_i$, which is
a function of the measured syndrome.
The decoding error for $C_1\ssi/C_2\ssi\mbox{}^{\perp}$ occurs only if
$e_i$ is outside $\tilde{\crI}=J+C_2\ssi\mbox{}^{\perp}$,
where $C_1\ssi$ is $J$-correcting. 
At this stage, the received word $y$ can be changed into
the interim estimate
\[
y'=y-\hat{e}=x+(e-\hat{e}).
\]

We employ bounded distance decoding here for simplicity,
though other schemes for classical concatenated codes, 
such as generalized minimum distance (GMD) decoding~\cite{forney}, 
are also applicable.
Then, the error $e$ is correctable 
if $e$ is such that the number of inner codes with erroneous decoding 
(the number of $i$ with $e_i \ne \hat{e}_i$)
is less than $\vart$, 
where we assume
the outer code $D_1$ is $\vart$-error-correcting.

The decoding for the outer code should be done based on
the latter half of the syndrome that comes from
the lower half of the parity check matrix in (\ref{eq:H4cat}).
This is possible as will be argued in Section~\ref{ss:synd_dec_cccss}.

\section{Syndrome Decoding for Concatenated Conjugate Codes \label{ss:synd_dec}}

\subsection{Preliminaries on Codes over Extension Fields \label{ss:ef}}

If $\bsb=(\beta_j)_{j=1}^{k}$ is a basis of $\myF$-linear vector
space $\myFk$, any element $\xi\in\myFk$ can be written as
\[
\xi= x_1 \beta_1 + \cdots + x_k \beta_k.
\]
The numerical row
vector $(x_1,\dots,x_{\degF})$ obtained in this way is 
denoted by $\varphi_{\bsb}(\xi)$.
Our arguments to be given 
rely on the next lemma.
\begin{lemma}\label{lem:expansion}
There exists a triple $(\varphi, \varphi',\MatF)$ that
consists of three bijections $\varphi:\myFk\to\myFpower{k}$,
$\varphi':\myFk\to\myFpower{k}$, and
$\MatF:\myFk\to\myFpower{k\times k}$ (the set of $k\times k$ matrices over
$\myF$)
with the following properties.
(i) $\varphi=\varphi_{\bsb}$ and  $\varphi'=\varphi_{\bsb'}$ 
for dual bases $\bsb$ and $\bsb'$.
(ii) We have
\[ 
\MatF(\xi) \varphi(\xi')\transp  = \varphi(\xi\xi')\transp,\quad
\varphi'(\xi) \MatF(\xi')  = \varphi'(\xi\xi')
\] 
and 
\[
\MatF(\xi)\MatF(\xi')=\MatF(\xi\xi'),\,\,\,
\MatF(\xi)+\MatF(\xi')=\MatF(\xi+\xi')
\]
for any $\xi,\xi'\in\myFpower{\degF}$.
\end{lemma}

{\em Remark}\/. In fact, we can show the stronger statement that
whenever $\bsb$ and $\bsb'$ are dual bases, for some $\MatF$,
the condition (ii) of Lemma~\ref{lem:expansion} holds with
$\varphi=\varphi_{\bsb}$ and  $\varphi'=\varphi_{\bsb'}$. 

A proof of the lemma and its remark, together with concrete forms of
$(\varphi, \varphi',\MatF)$, is included in Appendix~\ref{app:proof_expansion}.
The fact in Lemma~\ref{lem:expansion}, 
with `$\varphi'(\xi) \MatF(\xi')  = \varphi'(\xi\xi')$'
absent, has often been used in implementing codes over extension fields. 

Suppose we have an $[N, \varKgen]$ linear code $D$ over $\myFk$.
This can be used as a $[kN, k\varKgen]$ linear code $D'$ over $\myF$
if we apply some $\myF$-linear map from $\myFk$ onto $\myFpower{k}$
to each symbol of $D$.
Then, what is the parity check matrix of $D'$?

Let $H$ be a 
parity check matrix of $D$.
We extend the domain of $\varphi$ [$\varphi'$] to $\myFkpower{M}$,
where $M$ is a positive integer, in the natural manner:
We apply $\varphi$ [$\varphi'$] to each symbol of a word $x\in\myFkpower{M}$,
and denote
the resulting $kM$-dimensional vector over $\myF$ by $\varphi(x)$
[$\varphi'(x)$]. 
Our problem is to find a matrix $H'$ such that
\[ 
xH\transp=\zrvb \leftrightarrow \varphi(x)H'\transp=\zrvb,
\] 
where $\zrvb$ is the zero vector.
This will be accomplished if we find a matrix $H'$ such that
\begin{equation}\label{eq:H}
\varphi(xH\transp)=\varphi(x)H'\transp, \quad x\in\myFkpower{N}.
\end{equation}
Let $H=[h_{ij}]$ with $h_{ij}\in\myFk$.
Then, (\ref{eq:H}) holds for the matrix $H'=[\MatF(h_{ij})]$
with $\MatF$ as in Lemma~\ref{lem:expansion}.
This is a direct consequent of the first equation of 
condition (ii) of Lemma~\ref{lem:expansion},
which can be rewritten as 
$\varphi(\xi') \MatF(\xi) \transp  = \varphi(\xi\xi')$.
In particular, we have, for $H'=[\MatF(h_{ij})]$,
\begin{equation}\label{eq:pchgmt1}
\varphi(D)=\{ y\in\myFpower{kN} \mid yH'\transp =\zrvb \}.
\end{equation}

This simple logic also works if the pair $(\MatF,\varphi)$ is
replaced by $(\MatF\transp,\varphi')$, where $\MatF\transp$
is defined by $\MatF\transp(\xi)=\MatF(\xi)\transp$, $\xi\in\myFk$,
since $(\MatF\transp,\varphi')$ has a property of the same form
$\MatF(\xi')\transp \varphi'(\xi)\transp  = \varphi'(\xi\xi')\transp$.
Hence, 
\begin{equation}\label{eq:H''}
\varphi'(xH\transp)=\varphi'(x)H''\transp, \quad x\in\myFkpower{N}
\end{equation}
where $H''=[\Phi(h_{ij})\transp]$.

\subsection{Syndromes of Concatenated Quotient Codes \label{ss:synd_dec_cccss}}

Recall we have fixed two bases 
$\bsb=(\beta_j)_{j=1}^{k}$ and
$\bsb'=(\beta'_j)_{j=1}^{k}$ that are dual to each other
in constructing concatenated codes.
%
Now we easily see
$G'_{j,i}$ in (\ref{eq:H4cat}) are obtained from a parity check matrix 
$H$ of $D_1$ as follows.
We can use the arguments in Sections~\ref{ss:ef}
putting $H'=[\MatF(h_{ji})]$ with $D=D_1$.
We replace each row $\eta=(\eta_1,\dots,\eta_{k\ssi})$ of
$\MatF(h_{ji})$ by
\[
\sum_{m=1}^{k\ssi} \eta_m g'_m\mbssi ,
\]
and set the resulting $k\ssi \times n\ssi$ matrix  
equal to $G'_{j,i}$, 
$i\in\intint{1}{N}$, $j\in\intint{1}{M}$.

With the parity check matrix in (\ref{eq:H4cat}) 
and $G'_{j,i}$ constructed as above,
the latter half of the syndrome is the same as
\[
\varphi(x) H'\transp
\]
by (\ref{eq:gg'}), where $\varphi=\varphi_{\bsb}$.
Hence, known procedures to estimate the error pattern from the syndrome
for $D_1$ 
can be used to decode $\embone(D_1)$.

Note also that the parity check matrix of 
$\embone(D_1)+\overline{C_2^{\perp}}$ thus obtained is
a generator matrix of its dual $\embtwo(D_1^{\perp})+\overline{C_1^{\perp}}$.
Since $L_1$ and $L_2$ have the same form, generator matrices
of them are obtained similarly.

\section{Performance of Concatenated Conjugate Codes \label{ss:performance}}

We evaluate the performance of concatenate conjugate codes used on additive memoryless channels,
employing the bounded distance decoding as in Section~\ref{ss:dec}
for simplicity. 
Though the resulting bound on the decoding error probability
apparently admits of improvement in exponents
by GMD decoding~\cite[Chapter~4]{forney}, 
we do not pursue optimization of attainable exponents
staying at the issue of establishing achievable rates.

We know the existence of the sequence of $[[n,k]]$ conjugate code pairs 
$(C_1,C_2)$ over $\myF$
whose decoding error probabilities, say, $P_1$ for $C_1/C_2^{\perp}$ 
and $P_2$ for the other, are bounded by
\begin{equation}\label{eq:bound_inner}
\Pein=\max\{P_1, P_2 \}\le a_n q^{-n E(\rcl)}.
\end{equation}
Here, 
\begin{equation}\label{eq:rcl_r}
\rcl = \frac{\rqt+1}{2}, \quad \mbox{where} \quad  \rqt=\frac{k}{n}
\end{equation} 
and $a_n$ is polynomial in $n$~\cite{hamada03s,hamada05qc}.
This bound is attained by codes such that $k_1=k_2$ \cite{hamada03s,hamada05qc,hamada06expl,hamada06itw}.
Note
(\ref{eq:rcl_r}) is a rewriting of (\ref{eq:css_k}) with $k_1=k_2$,
and $\rcl$ is the rate $k_1/n$ of $C_1$
when it is viewed as a classical code. 
The exponent $E(\rcl)$ can be understood as the random coding exponent
(or it may be whatever is attainable by conjugate codes, e.g., 
$\max\{ E_{\rm r}(P,\rcl), E_{\rm ex}(P,\rcl+o(1))\}$ in \cite[Theorem~4]{hamada05qc}, 
which can also be attained by codes in \cite{hamada03s,hamada06expl,hamada06itw}).

We use $(C_1,C_2)$ as above for inner codes, and
generalized Reed-Solomon codes for outer codes $D_1$ and $D_2$
of the same dimension $K_1$,
and evaluate the concatenation $(L_1,L_2)$ of $(C_1,C_2)$ and $(D_1,D_2)$ as 
described in Section~\ref{ss:ccc}.
We consider an asymptotic situation 
where both $N$ and $n$ go to $\infty$,
$\Rcl = K_1/N$ approaches a fixed rate $\Rcl^*$, and 
$\rcl$ approaches a rate $\rcl^*$.
The decoding error probability $\Pej$ of $L_j/L_{\bar{j}}$,
where $\bar{1}=2$ and $\bar{2}=1$,
is bounded by
\begin{eqnarray*}
\Pej &\le & \sum_{i=\vart}^N \chooses{N}{i} \Peinj^i (1-\Peinj)^{N-i}\\
 &\le & q^{\vart \log_{q} \Peinj + (N-\vart) \log_{q}(1-\Peinj) + N h(\vart/N) }
\end{eqnarray*}
where $h$ is the binary entropy function, and $\vart=\lfloor (N-K_1)/2 \rfloor+1$
(for the second inequality, see, e.g., \cite[p.~446]{roman}; slightly weaker bounds can be found in other books on information theory). 
Taking logarithms and dividing by $\Noa=nN$,
and noting (\ref{eq:bound_inner}),
we have
\begin{eqnarray*}
\frac{1}{\Noa} \log_q \Pej &\le& 
\frac{\vart}{N} \Big[-E(\rcl)+ \frac{\log_q a_n}{n} \Big]\\
&&\!\!\!\!\! \mbox{}+ \frac{1}{n}\frac{N-\vart}{N} \log_q(1-\Peinj) + \frac{1}{n} h(\vart/N)
\end{eqnarray*}
for $j=1,2$.
Hence, 
the decoding error probability $\Pe$ of the concatenated code pair 
$(L_1,L_2)$, 
which is defined by $\Pe=\max\{ \Peone , \Petwo \}$, 
satisfies
\[
\limsup_{\Noa\to\infty} - \frac{1}{\Noa} \log_q \Pe 
\ge \frac{1}{2}\max(1-\Rcl^*) E(\rcl^*).
\]
This attainable exponent is the same as that discovered 
by Forney~\cite[Chapter~4]{forney} except the maximization range to be
explained.
Converting the rates into those of quotient codes by (\ref{eq:rcl_r}), namely,
by $\rcl^*=(r+1)/2$ and $\Rcl^*=(R+1)/2$, we have the next theorem.

\begin{theorem}\label{th:performance}
Assume we have a sequence of $[[n,k]]$ conjugate codes attaining
an error exponent $E((1+\rqt)/2)$ as in (\ref{eq:bound_inner}).
Then, there exists a sequence of
$[[\Noa,\Koa]]$ conjugate code pairs $(L_1,L_2)$ of the following properties.
(i) The rate $\Koa/\Noa$ approaches
a fixed number $\Roa$. 
(ii) The decoding error probability $\Pe$
is bounded by 
\begin{equation*}\label{eq:exp_ccc}
\limsup_{\Noa\to\infty} - \frac{1}{\Noa} \log_q \Pe 
\ge \frac{1}{4}\max_{rR=\Roa}(1-R)E((1+r)/2)
\end{equation*}
where the maximum 
is taken over $\{ (r,R) \mid 0 \le r \le 1,  0 \le R \le 1,  rR=\Roa \}$.
(iii)  The code $L_1/L_2^{\perp}$ and $L_2/L_1^{\perp}$ are decodable
with algorithms of polynomial complexity.
\end{theorem}

The attainable exponent, $E_{\rm L}(\Roa)$, in the theorem
is positive
whenever $E(\Roa)$ is positive. (A way to draw the curve 
of $E_{\rm L}(\Roa)=\max_{rR=\Roa}(1-R)E_{\rm a}(r)$ from that of another 
function $E_{\rm a}(r)$ is given in \cite[Fig.~4.3]{forney}.)

Hence, the achievable rate obtained in \cite{hamada03s}, which 
follows from the exponential bound in the form (\ref{eq:bound_inner}),  
is achievable
by codes for which polynomial decoding algorithms exist. 
For the simplest case where $\dmn=2$,
this rate is written in the form $1-2h(p)$ with a noise parameter $p$,
which is the probability of flipping the bit if the assumed channel
is the binary symmetric channel (BSC);
In short, the achievability comes from that both $C_1$ and $C_2$
achieve the capacity of the BSC;
By (\ref{eq:rcl_r}) or $\rcl^*=(r+1)/2$, the rate $\rcl^*=1-h(p)$ 
is converted into $r=1-2h(p)$.

\section{Discussions and Remarks \label{ss:dis_rem}}

\subsection{Related Code Constructions}

A special choice of $(D_1,D_2)$ and $(C_1,C_2)$ in our code construction
recovers results in \cite{KasamiLin88,GrasslGB99}. Theorem~\ref{th:duals_ccc}
for $C_1\ssi=C_2\ssi=\myFpower{k}$, $n\ssi=k$, $i\in\intint{1}{N}$,
was observed in \cite{KasamiLin88}. 
If $D_1=D_2$ and it is a Reed-Solomon (RS) code in addition, 
our code construction gives
the so-called quantum RS code~\cite{GrasslGB99}.
In this case, the inner codes are the $[n,n]$ code, not a real code,
so that the resulting code of length $nN$ is not a real concatenated code.

Theorem~\ref{th:duals_ccc} restricted to the case where
$C_2\ssi=\myFpower{n\ssi}$ and $k=k_1\ssi$, $i\in\intint{1}{N}$, appeared
in \cite{ChenLX01}.

Concatenated quantum codes are sometimes treated in the literature (e.g., \cite{hamada02c} and references therein). 
However, 
the literature has been lacking
cryptographic (quotient) codes that allow efficient decoding and
achieve the rate $1-2h(p)$~\cite{hamada03s},
which has been the (at least, short-term) goal of 
this issue of conjugate, or CSS, codes (e.g., \cite{MacKayMM04}).

\subsection{Remarks on Decoding Complexity}

We would need to be careful if we were to argue on efficient decoding
of quantum codes. 
In the quantum theoretical setting, 
one natural measure of the complexity is the number of 
primitive unitary operations (quantum gates) needed in a decoding
process. This is not the concern of this paper.

We evaluated the decoding complexity of 
cryptographic (quotient) codes, 
which uses only classical information processing~\cite{wyner75}. 
We remark in known applications of quotient codes to quantum cryptography, 
we need quantum mechanical devices only for modulation~\cite{ShorPreskill00,hamada03s,hamada06s}.

\subsection{Constructibility}

Though we have emphasized the efficiency of decoding,
our method of concatenation is also effective for 
constructibility.
A polynomial construction of codes that
achieve the rate $r=1-2h(p)$ is given in \cite{hamada06expl,hamada06itw}.
The minimum distance of constructive 
concatenated conjugate codes obtained with 
our method is larger than those known~\cite{hamada06md,hamada06itw}.

We remark that our evaluations on the reliability of conjugate code pairs
$(L_1,L_2)$ has direct implications on the reliability 
of the CSS quantum codes specified, 
as in the footnote in Section~\ref{ss:intro},
by $(L_1,L_2)$, which are involved
with quantum mechanical operations:
The fidelity of the CSS code is lower-bounded by $1-\Peone-\Petwo$
(see, e.g., \cite{hamada06s,hamada05qc}).

\section{Summary and Concluding Remarks \label{ss:sum}}

We brought Forney's idea of concatenating codes
into our issue of constructing long conjugate codes. 
The main technical issue resolved
is to concatenate conjugate code pairs 
retaining the constraint $C_2^{\perp}\le C_1$.
It was shown that the so-called Shannon rate
 $1-2h(p)$ of CSS-code-based cryptographic codes
is achievable with codes that allow polynomial decoding. 
Furtherance would be found in \cite{hamada06expl,hamada06md,hamada06itw}.

\section*{Acknowledgment}

The author wishes to thank O.~Hirota and A.~Hosoya for encouragement.

\appendices

\section{Preliminaries on Extension Fields \label{app:proof_expansion}}

\subsection{Companion Matrix \label{ss:ff}}

We begin with treating the basis
$\bsa=(\alpha^{j-1})_{j=1}^{k}$ with $\alpha$ being a primitive element
of $\myFk$.
We also use the following alternative visual notation for $\varphi_{\bsb}$ 
in the case of $\bsb=\bsa$.
\[
\begin{matrix}
|\\ \xi\\| 
\end{matrix}
= \varphi_{\bsa}(\xi)\transp =
\begin{bmatrix}
\xi_0\\ \vdots \\ \xi_{\degF-1}
\end{bmatrix}.
\]

Let $g(x)=x^{\degF}-g_{\degF-1}x^{\degF-1}\cdots-g_1 x -g_0$
be the minimum polynomial of $\alpha$ over $\myF$.
The companion matrix of $g(x)$ is 
\[
T=\begin{bmatrix} 
&0_{n-1} & g_0 \\
&{\Large I_{n-1}} & \begin{matrix} g_1 \\ \vdots \\ g_{\degF-1} \end{matrix}
\end{bmatrix}
\]
where $0_{\degF-1}$ is the zero vector in $\myFpower{\degF-1}$,
and $I_{\degF-1}$ is the $(\degF-1) \times (\degF-1)$ identity matrix.
Note that
\begin{equation}\label{eq:Ta}
T 
= \begin{bmatrix} | & & | \\
                     \alpha^{1} & \cdots & \alpha^{\degF}\\
                      | & & | 
       \end{bmatrix}.  
\end{equation}
Then, 
we have
\begin{equation}\label{eq:Tb}
T \begin{matrix}
|\\ \alpha^i \\| 
\end{matrix}
= \begin{matrix} | \\
                     \alpha^{i+1}\\
                  |
       \end{matrix},
\quad i\in\intint{0}{\dmn^\degF -2}. 
\end{equation}

{\em Proof of (\ref{eq:Tb}).}\/
Let $\varphi_{\bsa}(\alpha^i)=(x_1,\dots,x_\degF)$.
Then,
\[
T \varphi_{\bsa}(\alpha^i)\transp = \sum_{j=1}^{\degF} x_j \bab{\alpha^{j}} 
\]
by (\ref{eq:Ta}). The right-hand side can be written as
$\sum_{j=1}^n x_j \varphi_{\bsa}(\alpha^j)\transp=
\varphi_{\bsa}(\sum_{j=1}^n x_j \alpha^j)\transp=
\varphi_{\bsa}(\alpha\sum_{j=1}^n x_j \alpha^{j-1})\transp=
\varphi_{\bsa}(\alpha\alpha^{i})\transp$, completing the proof.
\enproof

We list properties of $T$, all of which easily 
follow from (\ref{eq:Tb}). 
By repeated use of (\ref{eq:Tb}), we have
\begin{equation}\label{eq:T3}
T^i \bab{\alpha^j}
=\bab{\alpha^{i+j}}
\end{equation}
for $i,j\in\intint{0}{\dmn^\degF -2}$. 
This implies
\begin{equation}\label{eq:T}
T^i = \begin{bmatrix} | & & | \\
                     \alpha^i & \cdots & \alpha^{i+\degF-1}\\
                      | & & | 
       \end{bmatrix} , \quad i\in\intint{0}{\dmn^\degF -2} 
\end{equation}
and hence,
\begin{equation}\label{eq:T1}
T^iT^j=T^{i+j}
\end{equation} 
and
\begin{equation}\label{eq:T2}
T^i+T^j=T^{\intk}
\end{equation} 
with $\intk$ satisfying
$\alpha^i+\alpha^j=\alpha^{\intk}$.  

To sum up, the map defined by
\[
\MatFa: \alpha^i \mapsto T^i,\quad i\in\intint{0}{\dmn^\degF -2},
\]
and $\MatFa(0)=O_{\degF}$ (zero matrix) is an isomorphism by (\ref{eq:T1})
and (\ref{eq:T2}):
\begin{gather}
\MatFa(\xi)\MatFa(\xi')=\MatFa(\xi\xi'), \label{eq:iso1}\\
\MatFa(\xi)+\MatFa(\xi')=\MatFa(\xi+\xi').\label{eq:iso2}
\end{gather}
%
%
By (\ref{eq:T3}), for any $\xi,\xi'\in\myFpower{\degF}$,
\begin{equation}\label{eq:T3alt}
\MatFa(\xi) \varphi_{\bsa}(\xi')\transp = \varphi_{\bsa}(\xi\xi')\transp.
\end{equation}

\subsection{Dual Bases}

In what follows, $\trace$ will be abbreviated as ${\rm Tr}$.
Let $\bsb=(\beta_j)_{j=1}^{k}$ and
$\bsb'=(\beta'_j)_{j=1}^{k}$ be bases of $\myFk$ that are dual to each other.
Namely,
\[
\tracenos \beta_l\beta'_m = \delta_{lm}.
\]
Then, for $\xi\in\myFk$, we have~\cite{LidlNied}
\[
\varphi_{\bsb'}(\xi)=(\tracenos \beta_1 \xi, \dots, \tracenos \beta_k \xi).
\]
For example, let $\bsa'$ denote the dual basis of $\bsa$.
Then,
\begin{equation}
\varphi_{\bsa'}(\xi)=(\tracenos \xi, \tracenos \alpha \xi, \dots, 
\tracenos \alpha^{k-1} \xi ). \label{eq:dual_vec_rep}
\end{equation}
In particular, it follows 
\begin{equation}\label{eq:T3alt_dual}
\varphi_{\bsa'}(\xi) \MatFa(\xi')  = \varphi_{\bsa'}(\xi\xi')
\end{equation}
for any $\xi,\xi'\in\myFpower{\degF}$, which makes good dual properties
with (\ref{eq:T3alt}).

{\em Proof of (\ref{eq:T3alt_dual}).}\/
We have
\begin{eqnarray*}
\lefteqn{\varphi_{\bsa'}(\alpha^i)T }\\
&=& \tracenos \alpha^i (0,\ldots,0, g_0) \\
&&\mbox{}+ \tracenos \alpha^{i+1} (1,0,\ldots,0, g_1)
+\cdots\\
&&\mbox{}+ \tracenos \alpha^{i+k-1} (0,\ldots,0,1, g_{k-1}) \\
&=& (\tracenos \alpha^{i+1}, \dots, \tracenos \alpha^{i+k-1},x),
\end{eqnarray*}
where
\begin{eqnarray*}
x&=&\tracenos(\alpha^ig_0+\cdots + \alpha^{i+k-1}g_{k-1})\\
&=&\tracenos\alpha^i(g_0+\cdots + \alpha^{k-1}g_{k-1})\\
&=&\tracenos\alpha^{i+k}.
\end{eqnarray*}
Hence, 
\begin{equation}\label{eq:Tb_dual}
\varphi_{\bsa'}(\alpha^i) T =\varphi_{\bsa'}(\alpha^{i+1}),
\end{equation}
which is the basic property that parallels (\ref{eq:Tb}).
Applying (\ref{eq:Tb_dual}) repeatedly, we obtain (\ref{eq:T3alt_dual}).
\enproof

\subsection{Proof of Lemma~\protect\ref{lem:expansion}}

By (\ref{eq:iso1}),  (\ref{eq:iso2}),  
(\ref{eq:T3alt}) and (\ref{eq:T3alt_dual}),  
we have a triple $(\varphi, \varphi',\MatF)$
that satisfies the conditions of the lemma. 
These are $\varphi=\varphi_{\bsa}$, $\varphi'=\varphi_{\bsa'}$
and $\MatF=\MatFa$.

Other solutions are given in the next subsection.

\subsection{Change of Bases \label{ss:ch_bs}}

Note (\ref{eq:T3alt}) and (\ref{eq:T3alt_dual}) 
can be rewritten as
\[ 
[\Lambda^{-1} \MatFa(\xi) \Lambda][ \Lambda^{-1} \varphi_{\bsa}(\xi')\transp] = 
[\Lambda^{-1} \varphi_{\bsa}(\xi\xi')\transp]
\] 
and
\[ 
[\varphi_{\bsa'}(\xi) \Lambda][ \Lambda^{-1} \MatFa(\xi')  \Lambda ]= [\varphi_{\bsa'}(\xi\xi') \Lambda]
\] 
with an invertible matrix $\Lambda$.
These imply that condition (ii) of Lemma~\ref{lem:expansion} 
is also satisfied by
$(\varphi,\varphi',\MatF)$ with
\begin{gather}
\varphi(\xi)\transp=\Lambda^{-1}\varphi_{\bsa}(\xi)\transp, 
\quad \varphi'(\xi)=\varphi_{\bsa'}(\xi)\Lambda, \nonumber\\
\MatF(\xi)=\Lambda^{-1}\MatFa(\xi)\Lambda.\label{eq:ch_basis}
\end{gather}

One may wonder if this newly obtained triple $(\varphi,\varphi',\MatF)$ 
has a relation to $\varphi_{\bsb}$ and $\varphi_{\bsb'}$
associated with a generic pair of dual bases $(\bsb, \bsb')$.
It does as we will see below.

Let $\bsb=(\beta_j)_{j=1}^{k}$, $\bsb'=(\beta'_j)_{j=1}^{k}$.
Recall that $\bsa=(\alpha_{j}=\alpha^{j-1})_{j=1}^k$ and 
$\bsa'=(\alpha'_j)_{j=1}^k$ is its dual.
We relate $\bsb$ with $\bsa$ by  
\[
\beta_i = \sum_j \alpha_j \lambda_{ji}
\]
and $\bsb'$ with $\bsa'$ by  
\[
\beta'_i = \sum_j \alpha'_j \lambda'_{ji}.
\]
Then,
\[
\varphi_{\bsa}(\xi)\transp = \Lambda \varphi_{\bsb}(\xi)\transp,\quad
\varphi_{\bsa'}(\xi)\transp = \Lambda' \varphi_{\bsb'}(\xi)\transp,
\]
where $\Lambda=[\lambda_{ij}]$ and $\Lambda'=[\lambda'_{ij}]$.
To retain the duality condition $\tracenos \beta_l\beta'_m =\delta_{lm}$,
$\Lambda$ and $\Lambda'$ should
satisfy
\[
\Lambda\transp \Lambda' = I_{k}.
\]
Hence, $(\varphi,\varphi')$ in (\ref{eq:ch_basis}) is nothing but
$(\varphi_{\bsb},\varphi_{\bsb'})$.

We have also shown the remark to Lemma~\ref{lem:expansion}
since the choice of $\bsb$ is arbitrary in the above argument.

\end{document}